\documentstyle[prd,aps,preprint]{revtex}
\begin{document}
\draft
\tighten
\preprint{TPR--93--24 \hspace{100mm} hep-ex/9407003}

\title{Unbiased estimators for correlation measurements}

\author{H.\ C.\ Eggers}

\address{Department of Physics, McGill University, Montr\'eal H3A 2T8,
Canada \\ and \\
Institut f\"ur Theoretische Physik, Universit\"at Regensburg,\\
D--93040 Regensburg, Germany \\ }

\author{P.\ Lipa\cite{OAW}}

\address{ Department of Physics, University of Arizona, Tucson AZ 85721 }

\date{September 9, 1993}

\maketitle

\begin{abstract}
Higher order correlation measurements involve multiple event
averages which must run over unequal events to avoid
statistical bias. We derive correction formulas for small event
samples, where the bias is largest, and utilize the results
to achieve savings in CPU time consumption for the star integral.
Results from a simple model of correlations illustrate the utility
and importance of these corrections. Single-event correlation
measurements such as in galaxy distributions and envisaged at RHIC
must take great care to avoid this unnecessary pitfall.
\end{abstract}


\section{Introduction}
\label{sec:intro}

In the hope of obtaining new insights into the old problem of
soft interactions in high energy physics, there has been much
interest in multiparticle correlations in the last few
years, spurred by new theoretical perspectives and
a large amount of multiparticle data in hadronic and nuclear
collisions \cite{Bia86a,DeW93a}. While various Monte Carlo codes and
analytical models often yield very similar behavior in rapidity
and $p_\perp$ distributions, they predict widely differing
particle correlations. Experimentally measured correlations
are therefore becoming an important and severe test of such
theoretical models.

Experience has shown, however, that correlation measurements
require considerably more subtle and sophisticated
understanding of statistics than single-particle
quantities do, and there has been much improvisation in
methodology and interpretation of data.
A clean and consistent statistical basis for such methodology
has become a matter of urgency.

Recently, we have shown how, through the use of
the correlation integral, the measurement of multiparticle
correlations can be greatly improved, both
in conventional variables such as rapidity and azimuthal angle
\cite{Egg93a} and in terms of relative momenta used in pion
interferometry \cite{Egg93d}. By deriving all quantities from first
principles, our techniques, besides greatly improving the accuracy
of correlation measurements, permit for the first time the direct
measurement of cumulants. Moments, while easily
measured, contain lower-order correlations.
Cumulants, testing the actual correlations, are to be
preferred, but they are hard to implement for at least two
reasons: they contain a hidden statistical bias and are expensive in
terms of CPU time.

The mentioned bias is present in {\it all} correlation measurements;
it is large for small data samples and strong correlations while
becoming negligible for large samples and weak correlations.
Our analysis provides the framework for understanding
and dealing with this bias in any present or future data set.

Secondly, correlation integral algorithms, while much superior to
conventional methods, run at least as the square  of the event
multiplicity and the sample size $N_{\rm ev}$. In understanding this
bias, we point the way to huge reductions in computer time also.
Defining for inner event averages a ``reduced sample average''
containing only $A$ events, and correcting for the resulting bias,
we obtain, compared to full event mixing, savings of a factor
$N_{\rm ev}/A$ for the star integral.
For a typical case with $N_{\rm ev} = 10^5$ events
and $A=100$, the savings amount to a factor $1000$ over full event mixing.

Besides the bias under discussion, there clearly are other biases,
both statistical and systematic, which greatly influence
multiparticle correlations. Typical unwanted but often important
effects include the ``empty bin effect'' \cite{Lip91a} and
contamination by trivial sources of particle correlations such as
Dalitz decays and gamma conversion \cite{EMU01-92a} or the
misidentification of pieces of a single track as two (highly
correlated) particles \cite{Lip90a}. All these have been shown
to be capable of drowning other correlations in the background.
Eliminating such biases is therefore a {\it sine qua non} of
multiparticle correlations. We take here a simple model of such
correlations, the split track model \cite{Lip91b}, to illustrate both
the use of the reduced event average with bias correction and the
effect such contamination may have on correlation data.

In Section \ref{sec:uprod}, we first explain the use and
significance of unbiased estimators and find a general form for
unbiased estimators of products of densities.
We develop the general formalism in Section \ref{sec:corterms} and
apply these in Section \ref{sec:starc} to the star integrals.
An example of behavior of the star integral as applied to the
split track model is given in Section \ref{sec:splt},
followed by an outline of  steps needed to measure unbiased
correlations in truly small samples and a brief discussion of
corrections for other correlation methods. We conclude with some
comments on small samples and single-event measurements.
First results regarding unbiased estimators can
be found in Ref.\ \cite{Kra93}. More recently, this formalism has been
applied to the problem
of normalization in a fixed-bin context \cite{KS94}.

\section{Unbiased estimators for products of distributions}
\label{sec:uprod}

We briefly remind the reader of some basics of statistical theory.
Suppose we have a random variable $U$ which for a given trial
(or ``event'' in the parlance of high energy physics) takes on a
value $\hat U$. For a finite number of events $N_{\rm ev}$, the set of
values of $\hat U$ make up a  {\it sample}, for which the
{\it sample average} of $U$ can be found,
$\langle U \rangle_s \equiv \sum_e \hat U_e /  N_{\rm ev}$.
By carrying out an infinite number of trials (the {\it population}),
one can theoretically determine the ``true'' behavior $ \bar U$ of
the random variable. The {\it expectation value}
$E[U]$ of a quantity $U$ is the value found over an infinite number
of trials,
\begin{equation}\label{esa}
 \bar U = E[U] = \lim_{  N_{\rm ev}\to\infty}
                 \sum_{e=1}^{N_{\rm ev}} \hat U_e / N_{\rm ev} \,.
\end{equation}
An experimental sample invariably consists of a finite number of
events, so that $E[U]$ cannot be found directly. A large part of
statistics occupies itself with the question how the information
contained in a limited sample can be extrapolated to estimate
its true behavior over the whole population. Rather than taking the
limit $N_{\rm ev}\to\infty$, one imagines that there are $\cal N$
samples, each with $N_{\rm ev}$ events and a particular value of
$\langle U \rangle_s$ for each. These sample averages themselves
form a distribution, the {\it sampling distribution}.
For infinite $N_{\rm ev}$, the sampling distribution of course
narrows to a delta function centered on $ \bar U$, but for finite
$N_{\rm ev}$ the sampling distribution has a nonzero width
independent of the number of samples ${\cal N}$.

There is no way to ascertain where the $\langle U \rangle_s$ obtained
for {\it one} experimental sample will fall in this distribution,
i.e.\ one can never claim with certainty that
$\langle U \rangle_s =  \bar U$.
All that can be achieved is to make sure that, even for finite
$N_{\rm ev}$, the {\it sampling average} of the sampling distribution
\begin{equation}\label{esab}
\left\{U \right\} \equiv \lim_{{\cal N}\to\infty}
{1\over {\cal N}} \sum_s \langle U \rangle_s
\end{equation}
equals the true value $ \bar U$.
Surprisingly, this is not generally true: for finite $N_{\rm ev}$,
$\left\{U \right\}$ is not necessarily equal to $ \bar U$.
When it is not, $U$ is
termed a {\it biased estimator} of $ \bar U$, and one attempts to
find a corresponding {\it unbiased estimator} $e(\bar U)$ which
does fulfil the condition
\begin{equation}\label{esac}
\left\{e(\bar U)\right\} =  \bar U \qquad
\mbox{\rm for all finite $N_{\rm ev}$}.
\end{equation}
Note:
Here and throughout this paper, we use the shortened notation
$e(\bar U)$ to denote the unbiased estimator {\it for} the true
value $\bar U$, i.e.\ the $\bar U$ inside the brackets is
not the argument of $e$ but the {\it desired result}. The set of
$\hat U_e$ of the experimental sample make up the arguments of $e$,
which in full notation should be written as
$e_{\bar U} (\hat U_1,\hat U_2, \ldots, \hat U_{N_{\rm ev}})$.

For the case of multiparticle physics, the basic random variables
$U$ correspond to the one-particle inclusive density of event $e$,
\begin{equation}\label{ese}
\hat\rho_1^e (\bbox{x}) =
\sum_{i=1}^{N} \delta(\bbox{x}_1 - \bbox{X}_i^e) \,,
\end{equation}
where $\bbox{X}_i$ are the set of measured coordinates
of the $N$ particles of the event,
and the corresponding $q$-th order densities for the
event $e$,
\begin{equation}\label{esf}
\hat\rho_q^e(\bbox{x}_1,\ldots,\bbox{x}_q) =
\sum_{i_1 \neq i_2 \neq \ldots \neq i_q}^{N}
\delta(\bbox{x}_1 - \bbox{X}_{i_1}^e) \, \cdots \,
\delta(\bbox{x}_q - \bbox{X}_{i_q}^e)
\,.
\end{equation}
These yield the sample average
$\rho_q = \langle \hat\rho_q^e \rangle_s
        = \sum_e \hat\rho_q^e /   N_{\rm ev}$,
identified with the usual experimental inclusive density
\begin{equation}\label{esd}
\rho_q(\bbox{x}_1,\bbox{x}_2,\ldots,\bbox{x}_q)
\equiv
{1\over \sigma_I} {d^q \sigma_{\rm incl}
\over d\bbox{x}_1 \, d\bbox{x}_2 \, \ldots d\bbox{x}_q }
\end{equation}
which is normalized to the factorial moment of the event
multiplicity $N$,
$\int \rho_q = \langle N^{[q]}\rangle
             = \langle N(N-1)\cdots(N-q+1) \rangle$.

It has long been known that the inclusive density is an unbiased
estimator for the true value, $\left\{\rho_q\right\} = \bar\rho_q$,
and so little attention has been paid to the theory of estimators
in high energy physics. Unlike a single inclusive density, however,
a product of two or more densities is a biased estimator.
This we illustrate for the simple example of the product of two
single-particle densities $\rho_1(\bbox{x}_1) \rho_1(\bbox{x}_2)$
before considering the general case.
The sampling average of $\rho_1\rho_1$ is
\begin{eqnarray}
\label{esg}
\left\{ \rho_1 \rho_1 \right\}
&=&
\left\{   N_{\rm ev}^{-2}
\sum_{e_1, e_2} \hat\rho_1^{e_1} \hat\rho_1^{e_2} \right\}
\nonumber\\
&=&
\left\{   N_{\rm ev}^{-2}
\sum_{e_1 \neq e_2} \hat\rho_1^{e_1} \hat\rho_1^{e_2} \right\}
+ \left\{   N_{\rm ev}^{-2}
\sum_{e_1} \hat\rho_1^{e_1} \hat\rho_1^{e_1} \right\}
\,,
\end{eqnarray}
i.e.\ there are $N_{\rm ev}$ out of the total $N_{\rm ev}^2$ terms
in which the two $\hat\rho_1$'s refer to the same event and thus
effectively introduce a correlation.
Because the densities of different events are independent, the
sampling average of their product factorizes, yielding the
true inclusive densities,\footnote{
The true value $\bar\rho_q$ can be
written as the sampling average of either the sample-averaged density
or of the single-event density,
$
\bar\rho_q = \left\{\rho_q \right\} = \left\{ \hat\rho_q \right\}
$,
because Eq.\ (\ref{esac}) is valid for single-event ``samples''
$N_{\rm ev} = 1$ also.
}
\begin{equation}\label{esi}
\left\{ \hat\rho_1^{e_1} \hat\rho_1^{e_2} \right\}   =
\left\{ \hat\rho_1^{e_1} \right\}
\left\{ \hat\rho_1^{e_2} \right\}
= \bar\rho_1  \bar\rho_1
\qquad\qquad\mbox{\rm if\ } e_1 \neq e_2
\end{equation}
so that
\begin{equation}\label{esj}
\{ \rho_1 \rho_1 \} =
    (1 - N_{\rm ev}^{-1}) \bar\rho_1 \bar\rho_1
     +   N_{\rm ev}^{-1}
         \left\{\hat\rho_1^{e_1} \hat\rho_1^{e_1}\right\} \,,
\end{equation}
meaning that $\left\{\rho_1\rho_1\right\}$ is not equal to
$\bar\rho_1 \bar\rho_1$ and thus $\rho_1\rho_1$ is
a biased estimator for the latter.
The culprit is clearly the equal-event part in Eq.\ (\ref{esg}).
For just one available sample, the needed unbiased estimator
for the true value ${\bar\rho_1}{\bar\rho_1}$ is the
unequal-event sum
\begin{equation}\label{esl}
e(\bar\rho_1 \bar\rho_1)
= {1\over   N_{\rm ev}^{[2]} }
    \sum_{e_1 \neq e_2} \hat\rho_1^{e_1} \hat\rho_1^{e_2} \,.
\end{equation}
The above simple example generalizes to the following result:
{\it Given a product of $K$ inclusive densities of order
$q_1$, $q_2, \ldots, q_K$, respectively, the unbiased estimator
for the product of true values is given by }
\begin{equation}\label{esm}
e(\bar\rho_{q_1} \bar\rho_{q_2} \cdots \bar\rho_{q_K})  =
{1\over   N_{\rm ev}^{[K]}  }
\sum_{e_1 \neq e_2 \neq \ldots \neq e_K}
\hat\rho_{q_1}^{e_1}  \hat\rho_{q_2}^{e_2}
              \cdots  \hat\rho_{q_K}^{e_K}
\,;
\end{equation}
for example, the unbiased estimator for
${\bar\rho_2} {\bar\rho_1} {\bar\rho_1}$ will be given by
$\sum_{e_1 \neq e_2 \neq e_3}
\hat\rho_2^{e_1} \hat\rho_1^{e_2} \hat\rho_1^{e_3} /N_{\rm ev}^{[3]}$.
This equation is the most important point of our paper.
In the following sections,
we explore the consequences for various correlation measurements
of taking only unequal events in products of densities.

Products such as in Eq.\ (\ref{esm}) can be written in terms of event
mixing, a procedure used heuristically before to normalize correlation
measurements. From here on, we distinguish three different
kinds of event mixing: Denoting the first event average by the index
$a$ and subsequent averages by $b, c$,
{\it full event mixing} is given by running all indices
over the full sample with $N_{\rm ev}$ terms,
\begin{equation}\label{ctaa}
{1\over   N_{\rm ev}   } \sum_{a=1}^{  N_{\rm ev}}
{1\over   N_{\rm ev} -1} \sum_{\scriptstyle b=1
                  \atop \scriptstyle b \neq   a}^{  N_{\rm ev}}
{1\over   N_{\rm ev} -2} \sum_{\scriptstyle {c=1}
                  \atop \scriptstyle c \neq a,b}^{  N_{\rm ev}}
\cdots \;,
\end{equation}
the {\it reduced event average} runs the inner event averages
over $A$ events only,
\begin{equation}\label{ctab}
{1\over   N_{\rm ev}} \sum_{a=1}^{  N_{\rm ev}}
{1\over A} \sum_{b=a-A}^{a-1}
{1\over A-1} \sum_{\scriptstyle c=a-A \atop\scriptstyle c\neq b}^{a-1}
\cdots  \;,
\end{equation}
while {\it fake event mixing} selects randomly a track from each
of $N$ different events (where $N$ itself must follow a Poisson
distribution) and does the standard analysis on a sample of
such fake events \cite{Lip92a}.
While full event mixing is exact, it is feasible
only for small samples, so that in practice the reduced average or fake
event procedures are chosen. The latter is easy to understand
and implement for the normalization $\rho_1^q$, but hard to
implement for the cumulant expansions introduced below.
We shall concentrate therefore on using the reduced event average.

\section{Correction terms for $K$-fold products}
\label{sec:corterms}

Before going into the details of unbiased estimators for the
various correlation measurements in current use, we establish the
general framework for these corrections which will be applicable
for all occurrences of products of random variables. To simplify
notation, we write for the single-event inclusive densities
$\hat\rho_q$ the variables
$ \hat U,  \hat V,  \hat W, \ldots$ and
$\langle \hat U \rangle_s =   N_{\rm ev}^{-1} \sum_e  \hat U_e$
the sample averages; the desired true values are
$ \bar U,  \bar V$ etc.  As in Section \ref{sec:uprod},
the desired unbiased estimator for a given product is obtained when
the single factors come from different events, as then the sampling
average factorizes,
\begin{equation}\label{ctb}
\left\{  \hat U^{e_1}  \hat V^{e_2}  \hat W^{e_3} \cdots \right\}_{
         e_1{\neq} e_2 {\neq} e_3 {\neq} \ldots }
=  \left\{  \hat U^{e_1} \right\}
   \left\{  \hat V^{e_2} \right\}
   \left\{  \hat W^{e_3} \right\} \cdots
=  \bar U  \bar V  \bar W \cdots  \,,
\end{equation}
and, to make full use of all events in the sample, the
sums over all (unequal) events are introduced.
Products of experimentally measured inclusive
densities, on the other hand, have unrestricted sums,
so that it is necessary to expand the unequal-event sums in terms
of unrestricted ones. Writing the Kronecker delta
$\delta_{e_1 e_2}$ as $\delta_{12}$ for short,
$\delta_{123} \equiv \delta_{e_1 e_2} \delta_{e_2 e_3}$ and so on,
we have for the double sum
\begin{equation}\label{ctc}
\sum_{e_1 \neq e_2}
= \sum_{e_1 e_2} - \sum_{e_1 = e_2}
= \sum_{e_1 e_2} (1 - \delta_{12}) \,.
\end{equation}
i.e.\ the factor $(1 - \delta_{12})$ forces the unrestricted sum
to the unequal-event sum. In third order, the corresponding
combinatorics are
\begin{center}
\begin{tabular}{r|l}
$e_1 = e_2 \neq e_3$ & $\delta_{12}(1-\delta_{23})$ \\
$e_1 = e_3 \neq e_2$ & $\delta_{13}(1-\delta_{23})$ \\
$e_2 = e_3 \neq e_1$ & $\delta_{23}(1-\delta_{13})$ \\
$e_1 = e_2  =   e_3$ & $\delta_{123}$ \\
$e_1 \neq e_2 \neq e_3$ &
$1 - \delta_{12} - \delta_{13} - \delta_{23} + 2\delta_{123}$ \,, \\
\end{tabular}
\end{center}
where the last line is obtained from the previous ones
by requiring that all cases have to add up to 1, so that
\begin{equation}\label{ctd}
\sum_{e_1 \neq e_2 \neq e_3}
= \sum_{e_1 e_2 e_3}
[ 1 - \delta_{12} - \delta_{13} - \delta_{23} + 2\delta_{123} ] ,
\end{equation}
while in fourth order,
\begin{equation}\label{cte}
\sum_{e_1 \neq e_2 \neq e_3 \neq e_4 }
=
\sum_{e_1 e_2 e_3 e_4}
[ 1 - \sum_{(6)}\delta_{12} + 2 \sum_{(4)}\delta_{123}
\nonumber\\
+ \sum_{(3)}\delta_{12}\delta_{34} - 6\delta_{1234} ]
\,;
\end{equation}
the brackets under the sums indicating the number of
permutations to be taken.

These expansions are utilized as follows. Let $A$ be the number of
events over which an average is performed,
$\langle \hat U \rangle = \sum_e  \hat U^e / A$
(this differs from the full sample average $\langle \hat U \rangle_s$
when doing reduced event mixing). The unbiased
estimator for $\bar U \bar V$ is expanded in second order to
\begin{equation}\label{ctf}
e(\bar U \bar V) =
{1\over A^{[2]}}
\sum_{e_1 \neq e_2}  \hat U^{e_1}  \hat V^{e_2}
= {A^2 \over A^{[2]} }  \langle \hat U \rangle \langle \hat V \rangle
- {A   \over A^{[2]} }  \langle \hat U  \hat V \rangle \,,
\end{equation}
and with $A^2/A^{[2]} = 1 + 1/(A{-}1)$,
\begin{equation}\label{ctg}
e( \bar U \bar V)
= \langle \hat U  \rangle \langle \hat V  \rangle
     - {1\over A - 1} \kappa_2( \hat U, \hat V) \,,
\end{equation}
where
\begin{equation}\label{cth}
\kappa_2( \hat U, \hat V) \equiv
\langle \hat U  \hat V \rangle
- \langle \hat U \rangle \langle \hat V \rangle \,,
\end{equation}
i.e.\ we get a correction consisting of a second-order
correlation, suppressed by a factor $(A -1)$.
Using Eq.\ (\ref{ctd}) and expanding
$A^3/A^{[3]} = 1 + 3/(A{-}1) + 4/(A{-}1)^{[2]}$ etc.,
we get for the third order unbiased estimator
\begin{eqnarray}
\label{cti}
e( \bar U \bar V \bar W)
&=&
{1 \over A^{[3]} }   \sum_{e_1 \neq e_2 \neq e_3 }
 \hat U^{e_1}  \hat V^{e_2}  \hat W^{e_3}
\nonumber\\
&=&
\langle \hat U \rangle  \langle \hat V \rangle  \langle \hat W \rangle
- {1\over A-1} \sum_{(3)}
\kappa_2( \hat U, \hat V)\, \langle \hat W \rangle
+ {2\over (A-1)^{[2]}}
\kappa_3( \hat U, \hat V, \hat W)  \,,
\end{eqnarray}
where
\begin{equation}\label{ctj}
\kappa_3( \hat U, \hat V, \hat W) \equiv
  \langle \hat U  \hat V  \hat W \rangle
- \langle \hat U  \hat V \rangle  \langle \hat W \rangle
- \langle \hat W  \hat U \rangle  \langle \hat V \rangle
- \langle \hat V  \hat W \rangle  \langle \hat U \rangle
+ 2 \langle \hat U \rangle \langle \hat V \rangle
    \langle \hat W \rangle
\end{equation}
is a third-order correlation, suppressed in Eq.\ (\ref{cti})
by a factor $1/(A-1)^{[2]}$. In fourth order, with
\begin{eqnarray}
\label{ctk}
\kappa_4( \hat U, \hat V, \hat W, \hat X)
&\equiv&
\langle \hat U  \hat V  \hat W  \hat X \rangle
- \sum_{(4)}  \langle \hat U  \hat V  \hat W \rangle
              \langle \hat X \rangle
\nonumber\\
&-&\mbox{} \sum_{(3)}  \langle \hat U  \hat V \rangle
                       \langle \hat W  \hat X \rangle
+ 2 \sum_{(6)}  \langle \hat U  \hat V \rangle
                \langle \hat W \rangle \langle \hat X \rangle
- 6 \langle \hat U \rangle \langle \hat V \rangle
    \langle \hat W \rangle \langle \hat X \rangle  \,,
\end{eqnarray}
we have
\begin{eqnarray}
\label{ctl}
e( \bar U \bar V \bar W \bar X)
&=&
\langle \hat U \rangle \langle \hat V \rangle
\langle \hat W \rangle \langle \hat X \rangle
     \nonumber\\
&-&\mbox{}
   {1\over A-1} \sum_{(6)} \kappa_2( \hat U, \hat V)
                \langle \hat W \rangle  \langle \hat X \rangle
     \nonumber\\
&+&\mbox{}
  {1\over (A-1)^{[2]}} \left(
     2 \sum_{(4)} \kappa_3( \hat U, \hat V, \hat W)
                  \langle \hat X \rangle
     + \sum_{(3)} \kappa_2( \hat U, \hat V)
                  \kappa_2( \hat W, \hat X)
                      \right)
     \nonumber\\
&-&\mbox{}
  {1 \over (A-1)^{[3]}} \left(
      6 \kappa_4( \hat U, \hat V, \hat W, \hat X)
    + 3 \sum_{(3)} \kappa_2( \hat U, \hat V)
                   \kappa_2( \hat W, \hat X)
                 \right)
\,.
\end{eqnarray}

\section{Bias corrections for the star integral}
\label{sec:starc}

As stressed previously, the quantity underlying all correlation
measurements is the inclusive density $\rho_q$:
Bose-Einstein measurements \cite{Egg93d},
fixed-bin factorial moments \cite{Bia86a} and cumulants \cite{Car90d},
as well as correlation integrals \cite{Egg93a} all sample $\rho_q$ in
the form of (unnormalized) factorial moments
\begin{equation}\label{esn}
\xi_q(\Omega) = \int_\Omega  d\bbox{x}_1 \, d\bbox{x}_2
              \ldots d\bbox{x}_q \,
\rho_q(\bbox{x}_1,\bbox{x}_2,\ldots,\bbox{x}_q)   \,.
\end{equation}
The only difference between these different correlation measurements
lies in the different choice of integration domain $\Omega$.

To explore the utility of unbiased estimators, let us look at the
so-called star integral, a particular method for
measuring multiparticle correlations \cite{Egg93a}.
The domain $\Omega$ for the star integral is given by the sum
of all spheres of radius $\epsilon$ centered around each of the
$N$ particles in the event.\footnote{
When a particle is closer than $\epsilon$ to the overall domain
boundaries, the sphere around it is truncated by the latter, so
that this definition is rigorous only for an infinite domain.
Boundary effects are, of course, the scourge of many correlation
measurements, even in astronomy \cite{Col92a}. Eq.\ (\ref{abb})
is rigorous for all domain sizes.}
The number of particles (``sphere count'')
within each of these spheres is, not counting the particle at
the center $\bbox{X}_{i_1}$,
\begin{equation}\label{ciad}
\hat n(\bbox{X}_{i_1},\epsilon)
\equiv
\sum_{i_2=1}^N \Theta(\epsilon - | \bbox{X}_{i_1} - \bbox{X}_{i_2} | )
\,,\ \ \ \ \ \ i_2 \neq i_1 \,,
\end{equation}
and the factorial moment of order $q$ is
\begin{equation}\label{ciac}
\xi_q^{\rm star}(\epsilon) =
 \left\langle \sum_{i_1} \hat n(\bbox{X}_{i_1},\epsilon )^{[q-1]}
 \right\rangle_s  .
\end{equation}
This can be derived rigorously \cite{Egg93a} from Eq.\ (\ref{esf})
using for $\Omega$ the equivalent definition
\begin{equation}\label{abb}
\xi_q^{\rm star}(\epsilon) =
\int \rho_q(\bbox{x}_1,\ldots,\bbox{x}_q) \,
\Theta_{12}\Theta_{13}\ldots\Theta_{1q} \,
d\bbox{x}_1 \ldots d\bbox{x}_q  \,,
\end{equation}
with the theta functions
$\Theta_{1j} \equiv \Theta(\epsilon - |\bbox{x}_1 - \bbox{x}_j|)$
restricting all $q{-}1$ coordinates $\bbox{x}_j$ to within a distance
$\epsilon$ of $\bbox{x}_1$.

For various reasons, it has become customary in high energy physics to
measure {\it normalized factorial moments} \cite{Bia86a}. Dividing by
the integral of the uncorrelated background $\rho_1^q$
over the same domain, the normalized star integral factorial moment is
\begin{equation}\label{cigc}
F_q^{\rm star}(\epsilon) \equiv
{\xi_q^{\rm star} \over \xi_q^{\rm norm}  }  =
{
\int \rho_q(\bbox{x}_1,\ldots,\bbox{x}_q) \,
\Theta_{12}\Theta_{13}\ldots\Theta_{1q} \,
d\bbox{x}_1 \ldots d\bbox{x}_q
\over
\int \rho_1(\bbox{x}_1)\ldots\rho_1(\bbox{x}_q) \,
\Theta_{12}\Theta_{13}\ldots\Theta_{1q} \,
d\bbox{x}_1 \ldots d\bbox{x}_q
} \,,
\end{equation}
where the denominator $\xi_q^{\rm norm}$ is given by the double
event average
\begin{equation}\label{evd}
\xi_q^{\rm norm}(\epsilon) =
 \left\langle \sum_{i_1}
 \left\langle \sum_{i_2}
\Theta(\epsilon - X_{i_1 i_2}^{ab})
 \right\rangle^{q-1}
 \right\rangle_s
\equiv
 \left\langle
\sum_{i_1}  \left\langle \hat n_b(\bbox{X}_{i_1}^a,\epsilon)
 \right\rangle^{q-1}
 \right\rangle_s ,
\end{equation}
with  $X_{i_1 i_2}^{ab} \equiv
| \bbox{X}_{i_1}^{a} - \bbox{X}_{i_2}^{b} |$ measuring the distance
between two particles taken from different events $a$ and $b$. The
(full $N_{\rm ev}$) outer event average and sum over $i_1$ are taken
over the center particle taken from event $a$, each of which is
used as the center of sphere counts
$\hat n_b(\bbox{X}_{i_1}^a,\epsilon)$
taken over all events $b$ in the (reduced) inner event average.

Having defined our terms, let us now analyse them from the point of
view of estimators. Because $\rho_q$ is an unbiased estimator
for the true $\bar\rho_q$, the numerator $\xi_q^{\rm star}$
is also unbiased and does not need correction. The denominator
$\xi_q^{\rm norm}$, however, is an integral over the biased estimator
$\rho_1(\bbox{x}_1)\cdots \rho_1(\bbox{x}_q)$. To shorten notation, we
abbreviate the sphere counts introduced previously by
\begin{eqnarray}
\label{cuja}
a &\equiv& \sum_{j}\Theta(\epsilon - X_{ij}^{aa})
   = \hat n(\bbox{X}_i^a,\epsilon), \ \ \ \  j\neq i \\
\label{cuj}
b &\equiv& \sum_j \Theta(\epsilon - X_{ij}^{ab})
   = \hat n_b(\bbox{X}_i^a,\epsilon) \,,
\end{eqnarray}
so that the uncorrected normalization is
$\xi_q^{\rm norm} = \left\langle\sum_i \langle b \rangle^{q-1}
\right\rangle_s$ for short.
The term inside the outer event average we write as
$\hat\xi_q = \langle b \rangle^{q-1}$, a $(q{-}1)$-fold product.
Inserting these $b$'s into Eqs.\
(\ref{ctg}), (\ref{cti}) and (\ref{ctl}),
unbiased estimators for the normalization moments are
found to be\footnote{
To avoid unnecessarily complicated notation, we omit here and
below the bar over ``hatted'' quantities inside the brackets.}
\begin{eqnarray}
\label{uej}
e(\hat\xi_2^{\rm norm}) &=&  \langle b  \rangle \,, \\
\label{ueja}
e(\hat\xi_3^{\rm norm})
&=&  \langle b  \rangle^2 - {\kappa_2(b,b) \over (A-1)}   \,, \\
e(\hat\xi_4^{\rm norm})
&=&  \langle b  \rangle^3
- {3  \langle b  \rangle \kappa_2(b,b) \over (A-1)}
     + {2\kappa_3(b,b,b) \over (A-1)^{[2]} } \,,
\\
e(\hat\xi_5^{\rm norm})
&=&  \langle b  \rangle^4
   - {6  \langle b  \rangle^2 \kappa_2(b,b) \over (A-1)   }
   + {  8  \langle b  \rangle \kappa_3(b,b,b)  +  3 \kappa_2^2(b,b)
       \over  (A-1)^{[2]}  } \nonumber\\
&&\mbox{} \quad\quad
   - { 6\kappa_4(b,b,b,b) + 9 \kappa_2^2(b,b)  \over (A-1)^{[3]}  }
\,,
\end{eqnarray}
where the definitions of $\kappa_q$ are given in Eqs.\
(\ref{cth}), (\ref{ctj}) and (\ref{ctk}).
In other words, the naive normalization $ \langle b  \rangle^{q-1}$
is corrected by correlations of order $q{-}1$ and lower,
suppressed by powers of $A$. The sample-averaged unbiased estimator
for the normalization is then
$e(\overline{\xi_q^{\rm norm}})
= \langle \sum_i e(\hat\xi_q^{\rm norm}) \rangle_s$,
and the bias-corrected normalized star integral is the ratio
\begin{equation}
\label{inb}
e(\bar F_q) = { \xi_q^{\rm star}
\over e(\overline{\xi_q^{\rm norm}})  } \,.
\end{equation}


A further possible bias must be tested, and, if necessary,
corrected for. Both numerator $\xi_q^{\rm star}$ and normalization use
the same sample, and thus will also contain a residual correlation
by referring to the same event during their respective averages.
The most obvious (but probably not the most elegant) way to remove this
correlation is to demand that the
denominator explicitly exclude each event $a$ currently under
consideration in the numerator.  The bottom-line unbiased
estimator for the normalized moment is therefore
\begin{equation}\label{inf}
e(\bar F_q) \equiv {1\over N_{\rm ev}} \sum_{a=1}^{N_{\rm ev}}
{ \hat\xi_q^a \over \hat D_q^a } \,.
\end{equation}
where $\hat D_q^a$ must now be
found from a product of $q$ single-particle densities restricted
additionally by the condition that all sums must exclude event $a$.

Consigning the details to the appendix, we here merely state the
results. Defining the ``correction function'' $\hat g_q^a$
implicitly by
\begin{equation}
\label{ing}
\hat D_q^a \equiv  e(\overline{\xi_q^{\rm norm}})
\left[ 1 - { \hat g_q^a \over N_{\rm ev} - 1} \right] ,
\end{equation}
the corrected normalized moment can be written as a geometric series
\begin{equation}
\label{ini}
e(\bar F_q)
= {1\over N_{\rm ev}}\sum_{a=1}^{N_{\rm ev}}
\left[ {\hat\xi_q^a \over e(\overline{\xi_q^{\rm norm}}) }
       \sum_{p=0}^\infty \left( \hat g_q^a \over N_{\rm ev}-1 \right)^p
\right]  .
\end{equation}
Suppressed by powers of $(N_{\rm ev}{-}1)$, this series converges
rapidly except for very small values of $N_{\rm ev}$. This means that
the correction due to correlation between numerator and denominator
can probably be neglected and only the $p{=}0$ term corresponding to
Eq.\ (\ref{inb}) need be kept. Should doubt arise as to the importance of
this correlation, the $p{=}1$ term
$\langle \hat\xi_q^a \hat g_q^a \rangle_s
  / (N_{\rm ev}{-}1)\,e(\overline{\xi_a^{\rm norm}})$
should be evaluated for the sample in question and compared to the
lowest-order term.


{\it Cumulants} are combinations of correlation functions constructed
in such a way as to become zero whenever any one or more of the
points $\bbox{x}$ becomes statistically independent of the others.
This is done so as to strip away the combinatorial background from
the correlation measurements,
\begin{eqnarray}
\label{cua}
C_2(\bbox{x}_1,\bbox{x}_2) &=& \rho_2(\bbox{x}_1,\bbox{x}_2) -
\rho_1(\bbox{x}_1)\rho_1(\bbox{x}_2) \,, \\
\label{cub}
C_3(\bbox{x}_1,\bbox{x}_2,\bbox{x}_3)
&=&\rho_3(\bbox{x}_1,\bbox{x}_2,\bbox{x}_3)
         -\ \rho_1(\bbox{x}_1)\rho_2(\bbox{x}_2,\bbox{x}_3)\nonumber\\
&&\mbox{} -\ \rho_1(\bbox{x}_2)\rho_2(\bbox{x}_3,\bbox{x}_1)
         -\ \rho_1(\bbox{x}_3)\rho_2(\bbox{x}_1,\bbox{x}_2)\nonumber\\
&&\mbox{} +\ 2\,\rho_1(\bbox{x}_1)\rho_1(\bbox{x}_2)\rho_1(\bbox{x}_3)
\,,
\end{eqnarray}
etc. Using combinations of conventional moments, they have been
measured for various experiments \cite{Car90d,EMU01-93a,NA22-93a}.
Integrating to get the unnormalized star integral cumulants
$ f_q$, defined by
\begin{equation}\label{cui}
 f_q(\epsilon) \equiv
\int C_q(\bbox{x}_1,\ldots,\bbox{x}_q) \,
\Theta_{12}\Theta_{13}\ldots \Theta_{1q} \,
d\bbox{x}_1\ldots d\bbox{x}_q  \,,
\end{equation}
we obtained previously \cite{Egg93a},
with $ f_q =  \langle \sum_i \hat f_q(i)  \rangle_s$,
\begin{eqnarray}
\label{cupa}
\hat f_2 &=& a - \langle b \rangle \,, \\
\hat f_3 &=& a^{[2]} -  \langle b^{[2]}  \rangle
              - 2 a \langle b \rangle + 2 \langle b \rangle^2
\,,
\end{eqnarray}
and so on for higher orders (see below). The second order $\hat f_2$
has only a single event average and so is unbiased,
$e(\hat f_2) = \hat f_2$. Correcting
according to Section \ref{sec:corterms} the last term
for $\hat f_3$ which involves a double event average,
the unbiased version becomes (again omitting the bars)
\begin{equation}\label{cupb}
e(\hat f_3)
=   a^{[2]} -             \langle b^{[2]}  \rangle
            - 2 a         \langle b        \rangle
            + 2           \langle b        \rangle^2
            - {2\over A-1} \kappa_2(b,b)  \,;
\end{equation}
similarly, the unbiased estimators for $\hat f_4$ and $\hat f_5$
are found to be
\begin{eqnarray}
e(\hat f_4)
&=& a^{[3]} -             \langle b^{[3]}  \rangle
            - 3 a^{[2]}   \langle b        \rangle
            - 3 a         \langle b^{[2]}  \rangle
            + 6           \langle b        \rangle
                          \langle b^{[2]}  \rangle
            + 6 a         \langle b        \rangle^2
            - 6           \langle b        \rangle^3
\nonumber\\
& &\mbox{}
+ {6\over A-1} \left[  \left( 3  \langle b  \rangle - a \right)
                       \kappa_2(b,b)
                        - \kappa_2(b,b^{[2]})
             \right]
- {12\over (A-1)^{[2]} }\kappa_3(b,b,b)
\,, \\
e(\hat f_5)
&=& a^{[4]} -             \langle b^{[4]}  \rangle
            - 4 a^{[3]}   \langle b        \rangle
            - 4 a         \langle b^{[3]}  \rangle
   \nonumber\\
& &\mbox{}  -  6 a^{[2]}  \langle b^{[2]}  \rangle
            +  8          \langle b        \rangle
                          \langle b^{[3]}  \rangle
            + 12 a^{[2]}  \langle b        \rangle^2
            +  6          \langle b^{[2]}  \rangle
                          \langle b^{[2]}  \rangle
\nonumber\\
& &\mbox{}  + 24 a        \langle b        \rangle
                          \langle b^{[2]}  \rangle
            - 36          \langle b        \rangle^2
                          \langle b^{[2]}  \rangle
            - 24 a        \langle b        \rangle^3
            + 24          \langle b        \rangle^4
\nonumber\\
& &\mbox{}
- {2\over A-1} \left[ \kappa_2(b,b) \left(
                      6 a^{[2]}      - 18  \langle b^{[2]}  \rangle
                - 36 a  \langle b  \rangle  + 72  \langle b \rangle^2
                                  \right)
             \right.
\nonumber\\
& &\mbox{}   \left. \qquad\qquad\qquad
                   + 4 \kappa_2(b      ,b^{[3]})
                   + 3 \kappa_2(b^{[2]},b^{[2]})
                   + \left( 12a - 36  \langle b  \rangle
                        \right) \kappa_2(b,b^{[2]})
             \right]
\nonumber\\
& &\mbox{}
+ {24\over (A-1)^{[2]}}
           \left[ 3 \kappa_2^2(b,b)
                 + (8  \langle b  \rangle - 2a) \kappa_3(b,b,b)
                 - 3 \kappa_3(b,b,b^{[2]})
           \right]
\nonumber\\
& &\mbox{}
- {72\over (A-1)^{[3]}}
           \left[ 2 \kappa_4(b,b,b,b) + 3 \kappa_2^2(b,b)
           \right]
\,.
\end{eqnarray}
The normalized cumulants are estimated by
\begin{equation}\label{cuh}
e(\bar K_q^{\rm star}(\epsilon)) \equiv
{  e(\bar f_q) \over e(\overline{\xi_q^{\rm norm}}) }
\ = \ {  \langle \sum_i e(\hat f_q) \rangle_s
   \over e(\overline{\xi_q^{\rm norm}}) }
\end{equation}
and must therefore be corrected for bias in both numerator and
denominator. For cumulants, too, the residual correlation between
numerator and denominator can be tested and corrected for; as for the
moments, we expect this correction to be negligible.
See the appendix for details.


A second useful form for star moments and cumulants are the so-called
{\it differential} moments: Here, one defines not only a maximum
distance $\epsilon_t$ but a minimum also, $\epsilon_{t-1}$
($t$ can define a sequence of such distances).
For a given combination of $q{-}1$ particles around a center
particle at $\bbox{X}_{i_1}$, at least one of these must lie inside
the spherical shell bounded by radii $\epsilon_{t-1}$
and $\epsilon_t$, while the others are restricted only by the
maximum distance $\epsilon_t$.
This definition leads rigorously \cite{Egg93a} to the simple and
efficient
prescriptions for measurement of the normalized differential moments
and cumulants
\begin{eqnarray}
\label{ddj}
\Delta F_q(t)
&=& {
 \left\langle \sum_i  a_t^{[q-1]} - a_{t-1}^{[q-1]}  \right\rangle_s
\over
 \left\langle \sum_i
 \left\langle b_t      \right\rangle^{q-1}  -
 \left\langle b_{t-1}  \right\rangle^{q-1}   \right\rangle_s
} \,, \\
\label{ddk}
\Delta K_q(t)
&=& {
 \left\langle \sum_i  \hat f_q(i,\epsilon_t) -
                      \hat f_q(i,\epsilon_{t{-}1})  \right\rangle_s
\over
 \left\langle \sum_i
 \left\langle b_t      \right\rangle^{q-1}  -
 \left\langle b_{t-1}  \right\rangle^{q-1}   \right\rangle_s
} \,,
\end{eqnarray}
using the shorthand $a_t \equiv \hat n(\bbox{X}_{i_1},\epsilon_t)$
and $b_t \equiv \hat n_b(\bbox{X}_{i_1},\epsilon_t)$.
Unbiased estimators are found by correcting individual terms
as set out for moments and cumulants above.

\section{An example: the split track model}
\label{sec:splt}

To illustrate the effect of bias and the use of the reduced inner
event averages for the star integral, we make use of a simple but
effective model, invented previously \cite{Lip91b} to simulate the
effects of spurious correlations introduced by Dalitz decays,
gamma conversion and the mismatching of tracks by detectors
\cite{Lip90a}.

For each ``event'', the {\it split track model\/} generates $P$
``points'' distributed uniformly inside a one-dimensional window,
with $P$ itself following a poisson distribution.
Each of these $P$ points is then either with probability $g$ split up
into $k$ ``particles'', all situated at exactly the same position,
or with probability $(1{-}g)$ becomes a single ``particle''.
The average multiplicity is thus
$\langle N \rangle = (1{-}g)\langle P\rangle  + gk\langle P\rangle $.
Clearly, the $k$ particles in a cluster are maximally correlated,
since they always fall within the same sphere,
no matter how small the radius $\epsilon$.

This simple model can be solved analytically and is known to yield
scaling cumulants $K_q$ for $q{\leq}k$, while cumulants of order greater
than $k$ are zero exactly \cite{Lip91b}.

We created $N_{\rm ev}=10,000$ events with average total number of
points $20$ and setting $g=0.1$ and $k=3$. This translates to an
average total multiplicity $\langle N \rangle=24$. Doing the reduced
event averages for the inner
($b$-)event average, only $A=11$ events rather than the full
$N_{\rm ev}$ were used. This means a savings of CPU time of about a
factor 1000 compared to full event mixing. Since there
are only three particles per cluster, the true cumulants
of fourth and fifth order are zero exactly. Both second and
third order cumulants should be nonzero and scaling.

In Figure 1, we show the effect of bias corrections
on the factorial moments $F_q$ and cumulants $K_q$
(note the different $y$-scales, both linear!). $F_2$ and $K_2$
have no bias corrections; for the higher orders, the
difference grows with increasing order $q$ and smaller sphere radius
$\epsilon$. As expected, the unbiased $K_4$ and $K_5$ are zero to within
statistical errors, while the biased $K_4$ and $K_5$ rise strongly.
The rise is due entirely to the equal-event bias which is the subject
of this paper. $F_4$ and $F_5$ contain contributions from second- and
third-order correlations \cite{Car90d} and therefore are not zero.

Note also that the biased estimate
lies {\it below} the unbiased one for the moments, while for
the cumulants, it lies {\it above} the unbiased estimator.
The reason is that the $F_q$ are corrected only through the
normalization $\xi_q^{\rm norm}$, which in Eqs.\ (\ref{uej}){\it ff.}
are all seen to be corrected downwards; the numerator
$\xi_q^{\rm star}$ is unbiased. For $K_q$, on the other hand, both
the numerator and denominator require bias corrections.

The corresponding differential moments and cumulants are shown in
Figure 2. The most important feature is that only the data point
corresponding to the smallest $\epsilon$ contains the split
track contributions to $\Delta K_2$ and $\Delta K_3$.
This must be so because all three particles belonging to a given cluster
are by construction separated by zero distance.

Secondly, the difference between biased and unbiased differentials is
much smaller than for the corresponding integral quantities $F_q$ and $K_q$.
This is because in Eqs.\ (\ref{ddj})--(\ref{ddk})
the subtraction of terms
($\langle b_t \rangle^{q-1} - \langle b_{t-1} \rangle^{q-1}$ etc)
means that corresponding corrections also
largely cancel. The only exception is the smallest-$\epsilon$ bin
where there are no  terms $\hat f_q(i,\epsilon_{t-1})$
and $b_{t-1}$ to subtract, so that the
bias corrections for this data point remain uncancelled.

It may be tempting to use a large value $A$ for inner event
averages while neglecting bias corrections rather than implementing
them. That this is usually not helpful is shown by Figure 3, where
we have plotted the dependence of the (biased, uncorrected)
$K_5$ on the number of events $A$ taken for the inner event average.
Again, the ``true'' value is $K_5\equiv 0$.
Clearly, the resulting curves converge rather slowly to zero even for large
$A$. The unbiased $K_5$, however, are virtually indistinguishable for
all values of $A$ shown here, meaning that, for the present
parameters, even the smallest value $A=11$ is sufficient to obtain
good results if the bias corrections are implemented. The factor 10
in CPU time needed for the $A=101$ case shown is thus largely wasted.
The only remaining advantage of using a larger $A$ is that
statistical errors become smaller (but the mean value remains the same).

The curves shown here are for the one-dimensional model; for
higher dimensions, the effect of split tracks on the correlation
is much larger since the rise of the cumulants goes roughly like
$\epsilon^{-d}$, where $d$ is the dimension of the phase space.

At this point we also comment on the use of different random
number generators. As can be seen in Figures 1--3,
the fifth order cumulant $K_5$ and differential $\Delta K_5$
show some deviation from the theoretical value of zero for
small $\epsilon$. We have tested various available random
number generators with the split track model, using exactly
the same parameters quoted above. It turns out that the
different generators produce substantially different results
for $K_5$ at small $\epsilon$, with some deviating above zero,
others below, with varying sizes of error bars. The calculation
of cumulants in the split track model
is clearly a {\it very} sensitive test of the
quality of a random number generator, just as it has proven
itself in ferreting out statistical and systematic experimental
biases. A really good random number generator should
yield results for $K_5$ within the split track model
which are compatible with zero.\footnote{
For the present examples, the generator RAN4 from Ref.\
\cite{Pre89a} was used.}

We therefore recommend that, before any experimental measurements
of correlations are attempted and compared to so-called
``random'' number data, {\it all\/} random number generators
first be tested whether they produce truly zero cumulants of
higher orders. Only when they do can any further conclusions
as to correlations in the data be drawn.

\section{Very small samples}

When the number of events in the experimental sample becomes
very small, of the order of 100 or less, full event mixing
may become unavoidable. In this case, of course, it becomes
mandatory to avoid the equal-event bias, otherwise the measurement is
simply wrong. Because for small samples CPU time is not an issue, the
best and most transparent method is directly to implement
the full unequal-event estimator of Eq.\ (\ref{esm}) for all
products in cumulants and normalization.

If for higher $q$ it does become advantageous to avoid direct
implementation of unequal-event algorithms, our procedures can be
used in modified form as follows.

Whereas the above bias corrections assumed that events $b,c,\ldots$
were always unequal to the ``outer'' event $a$, full event
mixing must allow and correct for all possible
combinations of equal and unequal events. Therefore, the
simple procedure of using the expansions of $ \bar U \bar V\ldots$
in terms $\langle \hat U \hat V \rangle$ etc.\ of Section
\ref{sec:corterms} cannot be applied directly; rather, one
must start from first principles and apply the sum combinatorics
to {\it all} sums. For example, the unbiased estimator for the
second order normalization becomes, after rearrangement
\begin{eqnarray}
\label{trb}
e(\overline{\xi_2^{\rm norm}} )
&=&
{1\over  N_{\rm ev}^{[2]}}
\sum_{a\neq b} \sum_{i,j} \Theta(\epsilon - X_{ij}^{ab})
\nonumber\\
&=&
{1\over  N_{\rm ev}^2}
\sum_{a,b} \sum_{i,j} \Theta(\epsilon - X_{ij}^{ab}) \nonumber\\
&&\mbox{}
- {1\over   N_{\rm ev}^{[2]}}  \sum_a
\left(
     \sum_{i,j} \Theta(\epsilon - X_{ij}^{aa})
      - {1\over   N_{\rm ev}}
          \sum_b \sum_{i,j} \Theta(\epsilon - X_{ij}^{ab})
\right)
\nonumber\\
&=&
 \left\langle \sum_i  \langle b  \rangle_s  \right\rangle_s
- {1\over   N_{\rm ev}-1}
 \left\langle \sum_i [ a+1 -  \langle b  \rangle_s ]
 \right\rangle_s \,,
\end{eqnarray}
so that one can infer
\begin{equation}\label{trc}
e(\hat\xi_2^{\rm norm}) =  \left\langle b  \right\rangle_s
- {1\over   N_{\rm ev}-1} [ a+1 -  \langle b  \rangle_s ]  \,.
\end{equation}
The extra ``1'' stems from the fact that the $i,j$ sums are not
restricted to unequal particles, so that the count always includes
the center particle also. Unlike the reduced event mixing
case of Eq.\ (\ref{uej}), which run only over the $A$ events following
$a$, the event averages here are performed over all $N_{\rm ev}$
events, including the $a=b$ case.

Using similar first-principle combinatorics, we find
for the full event mixing cumulant
\begin{equation}
\label{tre}
e(\hat f_2)
= a -  \langle b  \rangle_s
+ {1\over N_{\rm ev}-1} [ a+1- \langle b  \rangle_s ] \,.
\end{equation}
Higher order normalizations and cumulants are derived analogously.

\section{Corrections for Bose-Einstein and other correlations}
\label{sec:becor}

The prescription that only unequal events be used is of course
true for any kind of correlation measurement. In the case of
Bose-Einstein correlations, most experimental measurements to
date are for second order only, where the double event average
in the normalization is found through fake event mixing. Very few
higher order measurements exist, and these are in the form
of moments rather than cumulants, so that the problem did not arise
either \cite{Jur89a}.

Recently, we have derived formulae for the direct measurement of
cumulants in Bose-Einstein correlations \cite{Egg93d}.
The particular definition used for the $q$-particle
relative four-momentum,
\begin{equation}\label{quq}
Q^2 \equiv \sum_{\alpha < \beta = 1}^q - (p_\alpha - p_\beta)^2
\,,
\end{equation}
while convenient because it is directly related to the
$q$-particle invariant mass $M^2 = (p_1 + \ldots + p_q)^2$,
does not allow for a factorization of the multiple sums as
was the case for the star integral. For this reason, there
is little sense in deriving corresponding correction formulas;
rather, one simply must enforce all event sums  to refer to unequal
events as in Eq.\ (\ref{esm}) and do the full $q$-times event
average (or the corresponding reduced version).

There is one choice of the $q$-particle four-momentum that does
allow for factorization of the sums as in the star integral, namely
\begin{equation}\label{qur}
Q^2 \equiv \sum_{i=2}^q  -(p_1-p_i)^2 \,.
\end{equation}
For this case, corresponding
correction formulae can be derived and the savings in CPU
time achieved. It is unclear, however, whether such choice
of variable is preferable to the original choice of Eq.\ (\ref{quq})
for reasons other than convenience.

The situation is quite different for the traditional
(bin-based) factorial moments
$F_q = (1/M)\sum_m  \langle n_m^q  \rangle /  \langle n_m  \rangle^q$
of Bia\l as and Peschanski \cite{Bia86a}
and their cumulants \cite{Car90d}.
Here, the multiple event averages must be corrected
in the same way as the star integral; for example
\begin{equation}\label{qus}
 \langle n_m  \rangle^2 \longrightarrow
 \langle n_m  \rangle^2
- { \langle n_m^2 \rangle - \langle n_m  \rangle^2 \over N_{\rm ev}-1}
\end{equation}
and so on for higher order normalizations and cumulants.
The inherent instability and large error bars found for these
moments and cumulants, however, make it doubtful that these
corrections will make a discernible difference.

\section{Conclusions}

The statistical bias arising through the need for multiple event
averages must be understood and corrected for.
We have shown how the theory of unbiased estimators leads to
correction formulas for the star integral, thereby making it
possible to run it under fast algorithms without loss in
accuracy. For the envisaged large data samples, this savings
in CPU time may prove the difference between viability and
impossibility of correlation measurements in future.

For truly small samples, the correction for this bias is not
a tool for faster analysis but constitutive for a correct
measurement. Typical small samples are found in cosmic ray
data and in galaxy correlations as well as the subdivision of
inclusive data samples into fixed-multiplicity subsamples.
All these must take cognizance of the bias and correct for it.

This brings us to the subject of single-event measurements:
event mixing is, of course, not possible when there is just one
available. For the proposed measurement of Bose-Einstein correlations
in single events in nuclear collisions at RHIC and LHC, the
solution is clearly to normalize by event mixing based on
a sample of similar events. Most notably, this mixing
sample should have the same multiplicity and general characteristics;
such requirements will necessarily restrict the sample to
relatively few events, so that the bias corrections may become
important.

Galaxy distributions, on the other hand, present a much more
difficult task: there is no pool of big bang events to
make up the uncorrelated background. So far, the preferred
solution was to assume a uniform distribution on a sufficiently
large scale. Recent results on the large-scale structure of
the universe, however, make this assumption increasingly
untenable. The only alternative route would appear to be to
select a number of windows in the sky (with about the same overall
galaxy count as the window used for the numerator) and,
neglecting the long-range correlations, count these as different
``events''. In this way, no assumption of overall uniformity
need be made.

\acknowledgements
We thank our colleagues of the Max-Planck-Institut in Munich
as well as P.\ Carruthers and B.\ Buschbeck
for stimulating discussions. Special thanks to B.\ Wosiek and
F.\ Botterweck for criticism  and checking the programs.
HCE thanks the Alexander von Humboldt Foundation for support.
This work was supported in part by the US Department of Energy, grant
DE--FG02--88ER40456, and by the Austrian Fonds zur F\"orderung der
wissenschaftlichen Forschung, Project No.\ P8259--TEC.

\appendix
\section*{Unbiased estimators for normalized moments}

In this appendix, we derive the correction functions $\hat g_q$
to be used for checking for residual correlations between numerator
and denominator of the normalized factorial moment and cumulant.
Our notation will be as follows:
we use roman letters $a,b,\ldots$ for the event indices of
the numerator of the normalized moments, and greek letters
$\alpha,\beta,\ldots$ for the denominator.

\subsection{Reduced event mixing}

The numerator of the biased uncorrected $F_2$ is given by
\begin{eqnarray}
\label{rac}
\xi_2 &=& {1\over N_{\rm ev}} \sum_{a=1}^{N_{\rm ev}} \hat\xi_2^a \,,
\\
\hat\xi_2^a &=& \sum_{i\neq j}\Theta(\epsilon - X_{ij}^{aa}) \,,
\end{eqnarray}
while the denominator is
\begin{equation}\label{rad}
e(\overline{\xi_2^{\rm norm}}) =
{1\over  N_{\rm ev} A } \sum_{\alpha=1}^{N_{\rm ev}}
\sum_{\beta=\alpha-A}^{\alpha-1} T_{\alpha\beta} \,,
\end{equation}
with
\begin{equation}\label{rae}
T_{\alpha\beta} = \sum_{i,j}\Theta(\epsilon - X_{ij}^{\alpha\beta}) \,.
\end{equation}
Note that the inner $\beta$-average $\langle T_{\alpha\beta} \rangle$
is equal to
$\sum_i \langle b \rangle$ in the shortened notation of Eq.\ (\ref{cuj}).

In order to get an unbiased estimator $e(\bar F_2)$ of the normalized
second order factorial moment, we exclude explicitly from the
denominator the event $a$ used in the numerator event sum:
\begin{equation}\label{raf}
e(\bar F_2) = {1\over N_{\rm ev}} \sum_{a=1}^{N_{\rm ev}}
{ \hat\xi_2^a \over \hat D_2^a }
\end{equation}
where the denominator is now $a$-dependent:
\begin{eqnarray}
\label{eq:rag}
\hat D_2^a
&=& {1\over (N_{\rm ev}-1)} \sum_{\alpha=1}^{N_{\rm ev}} (1-\delta_{a\alpha})
    {1\over A} \sum_{\beta} T_{\alpha\beta}
\nonumber\\
&=& {1\over (N_{\rm ev}-1)} \sum_{\alpha=1}^{N_{\rm ev}} (1-\delta_{a\alpha})
\left[
 {1\over A} \sum_{\beta=\alpha-A  }^{\alpha-1} (1-C_{a\alpha}) T_{\alpha\beta}
+{1\over A} \sum_{\beta=\alpha-A-1}^{\alpha-1}
  (1-\delta_{\beta a})C_{a\alpha}\,  T_{\alpha\beta}
\right] ,
\end{eqnarray}
and the condition
\begin{equation}\label{rah}
C_{a\alpha} = \sum_{u=a+1}^{a+A} \delta_{\alpha u}
\end{equation}
is unity whenever $\alpha$ is in the range
$[a{+}1,\ldots,a{+}A]$ and zero otherwise
(note that $C_{a\alpha} \delta_{a\alpha} = 0$). The reason for the splitting
of the $\beta$-sum is that whenever $\alpha$ is in this range, the index
$\beta$ must ``jump'' the $a$-event, meaning that
the count must start at $\alpha{-}A{-}1$. The form (\ref{eq:rag}) thus
explicitly excludes the currently-used numerator event $a$.

To find the relation between $e(\bar F_2)$ and $\bar F_2$, we
factor out of $\hat D_2^a$ the usual normalization and write the remainder
in terms of a function $\hat g_2^a$ which is to be determined:
\begin{equation}\label{rahf}
\hat D_2^a \equiv  e(\overline{\xi_2^{\rm norm}})
\left[ 1 - { \hat g_2^a \over N_{\rm ev} - 1} \right] .
\end{equation}
The moment estimator is then a geometric series
\begin{eqnarray}
\label{eq:rai}
e(\bar F_2)
&=& {1\over N_{\rm ev}}\sum_{a=1}^{N_{\rm ev}}
\left[ {\hat\xi_2^a \over e(\overline{\xi_2^{\rm norm}}) }
       \sum_{p=0}^\infty \left( \hat g_2^a \over N_{\rm ev}-1 \right)^p
\right]
\nonumber\\
&=& { \langle \hat\xi_2^a \rangle_s
\over e(\overline{\xi_2^{\rm norm}})  }
+ {1\over (N_{\rm ev}-1)}\;
{\left\langle \hat\xi_2^a \hat g_2^a \right\rangle_s
\over e(\overline{\xi_2^{\rm norm}}) }
+ \cdots
\end{eqnarray}
which usually converges rapidly. The correction function
$\hat g_2^a$ is found as follows. The quantity in the square brackets
of Eq.\ (\ref{eq:rag}) yields, on rearrangement,
\begin{equation}\label{rak}
  {C_{a\alpha} \over A}(T_{\alpha,\alpha-A-1} - T_{\alpha,a})
+  {1\over A}\sum_{\beta=\alpha-A}^{\alpha-1} T_{\alpha\beta} ,
\end{equation}
so that
\begin{equation}
\label{ral}
\hat D_2^a - e(\overline{\xi_2^{\rm norm}})
= \sum_{\alpha=1}^{N_{\rm ev}} {1\over A}
\left[
   { C_{a\alpha} ( T_{\alpha,\alpha-A-1} - T_{\alpha a} ) \over N_{\rm ev}-1}
   + \sum_{\beta=\alpha-A}^{\alpha-1}
     \left(
       { (1-\delta_{a\alpha}) T_{\alpha\beta} \over N_{\rm ev} - 1 }
                          - { T_{\alpha\beta} \over N_{\rm ev} }
     \right)
\right]  .
\end{equation}
After changing to index $u=\beta+A+1$, the $\delta_{\alpha a}$ term becomes
$\sum_{\alpha\beta} \delta_{a\alpha}T_{\alpha\beta}
= \sum_{u=a+1}^{a+A} T_{a,u-A-1}$, which yields,
using Eqs.\ (\ref{rad}) and (\ref{rah}),
\begin{equation}\label{ram}
\hat D_2^a - e(\overline{\xi_2^{\rm norm}}) =
{1\over N_{\rm ev}-1}
\left[
e(\overline{\xi_2^{\rm norm}})
- {1\over A} \sum_{u=a+1}^{a+A} ( T_{a,u-A-1} - T_{u,u-A-1} + T_{a,u} )
\right] ,
\end{equation}
so that we can identify
\begin{equation}\label{ran}
\hat g_2^a =  -1 +
\left[
{1\over e(\overline{\xi_2^{\rm norm}}) A} \sum_{u=a+1}^{a+A}
( T_{a,u-A-1} - T_{u,u-A-1} + T_{a,u} )
\right] .
\end{equation}
Implementing this type of correction thus involves keeping the
sphere counts of events mixed within a range $[a-A,\ldots,a+A]$.

\subsection{Full event mixing}

Correcting for bias in the case of full event mixing is somewhat easier
than for the reduced event mixing above because the $\beta$-sum
does not have to be split up. The equivalent definitions are
\begin{equation}\label{fua}
e(\overline{\xi_2^{\rm norm}}) =
{1\over N_{\rm ev}(N_{\rm ev}-1)}
\sum_{\alpha,\beta=1}^{N_{\rm ev}} (1-\delta_{\alpha\beta})
T_{\alpha\beta}
\end{equation}
and
\begin{equation}\label{fub}
\hat D_2^a = {1\over (N_{\rm ev}{-}1)(N_{\rm ev}{-}2)}
\sum_{\alpha,\beta =1}^{N_{\rm ev}}
(1-\delta_{a\alpha}) (1-\delta_{a\beta}) (1-\delta_{\alpha\beta})
T_{\alpha\beta} \,,
\end{equation}
giving, using the symmetry of $T_{\alpha\beta}$,
\begin{eqnarray}
\label{eq:fuc}
\hat D_2^a - e(\overline{\xi_2^{\rm norm}})
&=&
{1\over N_{\rm ev}^{[3]}}
\sum_{\alpha\beta} (1-\delta_{\alpha\beta}) T_{\alpha\beta}
[ N_{\rm ev}
(1 - \delta_{a\alpha} - \delta_{a\beta} + \delta_{a\alpha\beta})
  - (N_{\rm ev}-2) ]
\nonumber\\
&=&
{ 2 \over N_{\rm ev}-2 }
\left[
{1\over N_{\rm ev}^{[2]} }
\sum_{\alpha\beta} (1-\delta_{\alpha\beta}) T_{\alpha\beta}
- {1\over N_{\rm ev}-1} \sum_{\beta} (1-\delta_{a\beta}) T_{a\beta}
\right]
\nonumber\\
&=&
{ 2 \over N_{\rm ev}-2 }
\left( e(\overline{\xi_2^{\rm norm}})
- \sum_i \langle b \rangle_a \right) ,
\end{eqnarray}
the second term being an event mixing average performed around
tracks $i$ of (numerator) event $a$ only, and hence
\begin{equation}\label{fud}
\hat g_2^a =
2\left( { \sum_i \langle b \rangle_a
\over e(\overline{\xi_2^{\rm norm}}) } -1 \right)
\,.
\end{equation}
The difference between this and the reduced event mixing case is that
the former keeps the ``mixing tail'' to strictly $A$ events, so that
even for the maximal $A=N_{\rm ev}{-}2$ it always leaves out one event in
the mixing. The full event mixing outlined above, on the other hand,
changes from mixing with $N_{\rm ev}{-}2$ events in $\hat D_2^a$ to
$N_{\rm ev}{-}1$ in $e(\overline{\xi_2^{\rm norm}})$.
The power series expansion now reads
\begin{equation}\label{fue}
e(\bar F_2) = { \langle \hat\xi_2^a \rangle_s
\over e(\overline{\xi_2^{\rm norm}})  }
+ {1\over (N_{\rm ev}-2)}\;
{\langle \hat\xi_2^a \hat g_2^a \rangle_s
\over e(\overline{\xi_2^{\rm norm}}) }
+ \cdots
\end{equation}

\subsection{Corrections for higher order}

For higher orders, a similar prescription would be followed in
eliminating bias arising from numerator-denominator correlations.
The unbiased form is
\begin{equation}\label{haf}
e(\bar F_q) = {1\over N_{\rm ev}} \sum_{a=1}^{N_{\rm ev}}
{ \hat\xi_q^a \over \hat D_q^a } \,.
\end{equation}
With the understanding that all indices $\alpha_i$ are kept
strictly unequal to each other throughout,
\begin{equation}
\label{hag}
\hat D_q^a = {1\over (N_{\rm ev}-1)}
\sum_{\alpha_1=1 \atop \alpha_1 \neq a}^{N_{\rm ev}}
\left[
   {1\over A^{[q-1]}}
\sum_{\alpha_2,\ldots,\alpha_q \atop =\alpha_1 -A}^{\alpha_1 -1}
(1-C_{a\alpha_1})\, T_{\alpha_1,\ldots,\alpha_q}
+  {1\over A^{[q-1]}}
\sum_{\alpha_2,\ldots,\alpha_q \atop =\alpha_1 -A-1}^{\alpha_1 -1}
C_{a\alpha_1}    \, T_{\alpha_1,\ldots,\alpha_q}
\right]
\end{equation}
where
\begin{equation}\label{hae}
T_{\alpha_1,\alpha_2,\ldots,\alpha_q}
= \sum_{i_1,i_2,\ldots,i_q}
\Theta_{i_1 i_2}^{\alpha_1 \alpha_2}
\Theta_{i_1 i_3}^{\alpha_1 \alpha_3}
\ldots
\Theta_{i_1 i_q}^{\alpha_1 \alpha_q}
\end{equation}
with
$\Theta_{i_1 i_2}^{\alpha_1 \alpha_2} =
\Theta(\epsilon - X_{i_1 i_2}^{\alpha_1 \alpha_2})$ as usual.
Note that $T$ is symmetric in all indices except $\alpha_1$.
The idea is then to expand $\hat D_q^a$ in terms of the corresponding
$a$-independent normalization $e(\overline{\xi_q^{\rm norm}})$ and a
correction function $\hat g_q^a$. The unbiased normalized moment
is then given by an expansion of the form of Eq.\ (\ref{eq:rai})
in powers of $N_{\rm ev}{-}1$.

By excluding one event from the sum, we are explicitly breaking the
symmetry of the sphere counts that permitted factorization of the
multiple sums, so that the correction function $\hat g_q^a$
becomes rapidly more complex with $q$. Here, we merely outline
the results for third order.
We have from Eq.\ (\ref{hag}), after going through similar steps
as for $q=2$,
\begin{eqnarray}
\label{eq:hah}
\hat D_3^a - e(\overline{\xi_3^{\rm norm}}) &=&
{1\over N_{\rm ev} - 1}
\left[
e(\overline{\xi_3^{\rm norm}})
+ {1\over A^{[2]} } \sum_{u=a+1}^{a+A}
\left(
2\sum_{\beta=u-A}^{u-1} ( T_{u,u-A-1,\beta}  - T_{u,a,\beta} )
\right. \right.\nonumber\\
& &\mbox{} \left. \left.
- 2 T_{u,a,a}
- \sum_{\beta=a-A}^{a-1} (1-\delta_{\beta,u-A-1}) T_{a,u-A-1,\beta}
\right)
\right] .
\end{eqnarray}
Note that $e(\overline{\xi_3^{\rm norm}})$ itself is an unbiased
estimator obtained from the biased form through Eq.\ (\ref{ueja}).
The correction function is then
\begin{eqnarray}\label{hai}
\hat g_3^a = -1
&+& {1\over A^{[2]}} \sum_{u=a+1}^{a+A}
   \left[
2\sum_{\beta=u-A}^{u-1} ( T_{u,a,\beta} - T_{u,u-A-1,\beta} )
   \right. \nonumber\\
&-&\left.
    2 ( T_{u,a,a} - T_{u,a,u-A-1} )
+ \sum_{\beta=a-A}^{a-1} (1 - \delta_{\beta,u-A-1}) T_{a,u-A-1,\beta}
   \right]  .
\end{eqnarray}
Because the correction $\hat g_q^a$ applies to the denominator
only, cumulants of higher order are immediately found from
\begin{eqnarray}
\label{eq:haj}
e(\bar K_q)
&=& {1\over N_{\rm ev}} \sum_{\alpha=1}^{N_{\rm ev}}
{\sum_i e(\hat f_q^a) \over \hat D_q^a }
\nonumber\\
&=&
{ \langle \sum_i e(\hat f_q^a) \rangle_s
\over e(\overline{\xi_q^{\rm norm}})  }
+ {1\over (N_{\rm ev}-1)}\;
{\langle \hat g_q^a \sum_i e(\hat f_q^a) \rangle_s
\over e(\overline{\xi_q^{\rm norm}} )}
+ \cdots
\end{eqnarray}

For the case of full event mixing, we obtain for the moments
\begin{equation}\label{hak}
e(\bar F_q) = {\langle \xi_q^a \rangle_s
\over e(\overline{\xi_q^{\rm norm}})  }  +
{1\over (N_{\rm ev} - q)}
{\langle \hat\xi_q^a \hat g_q^a \rangle_s
\over e(\overline{\xi_q^{\rm norm}}) }
+ \cdots
\end{equation}
where for third order
\begin{equation}\label{hal}
\hat g_3^a = -3  +
{1\over (N_{\rm ev}-1)(N_{\rm ev}-2)}
\sum_{\alpha\neq\beta = 1\ (\neq a)}^{N_{\rm ev}}
( T_{a\alpha\beta} + 2 T_{\alpha a \beta} )  \,.
\end{equation}



\section*{List of Figures}

\vspace{10mm}

\noindent
Figure 1:
Normalized factorial moments $F_q$ and cumulants $K_q$ for
$q=2,\ldots,5$ for the split track model, with 10\% of the
points split up into 3 tracks. For the inner event average
to calculate the $\hat n$ sphere counts, only $A=11$ events
were used rather than the full event mixing of $N_{\rm ev}=10,000$
events (i.e.\ shortening the CPU time by  a factor $\sim 1000$).
The biased moments and cumulants are clearly wrong ($K_4$
and $K_5$ should be zero), while the unbiased version are fine.

\vspace{20mm}

Figure 2:
Differential moments and cumulants. As the three split tracks are all
at the same point, the correlation due to their presence is always
contained in the smallest bin; this is clearly visible as the
single point in the unbiased $\Delta K_2$ and $\Delta K_3$.

\vspace{20mm}

Figure 3:
Full event mixing (using all $N_{\rm ev}$ events for the inner event
averages) is not a useful alternative to bias corrections.
As shown here, one needs upwards of $A=101$ events in the inner
loop to make the biased estimate approach that of the true value
$K_5=0$; the unbiased estimators (filled circles)
of $K_5$, on the other hand, all lie close to zero even for $A=11$
so that this small number is sufficient for a good estimate.
CPU time is roughly proportional to $A$, i.e.\ a factor
9 larger for $A=101$ than for $A=11$.


\begin{thebibliography}{99}


\bibitem[*]{OAW}
Present address: Institut f\"ur Hoch\-en\-er\-gie\-phy\-sik
der \"Osterreichischen Akademie der Wissenschaften,
Ni\-kolsdorfergasse 18, A--1050 Vienna, Austria, \\
v2033dag@awiuni11.edvz.univie.ac.at

\bibitem{Bia86a}A.\ Bia\l as and R.\ Peschanski,
            Nucl.\ Phys.\ {\bf B273}, 703 (1986);
            {\bf B308}, 857 (1988).


\bibitem{DeW93a}For a review, see
           E.\ A.\ de Wolf, I.\ Dremin and W.\ Kittel,
           preprint HEN--362 (1993), to appear in Phys.\ Rep.

\bibitem{Egg93a}H.\ C.\ Eggers, P.\ Lipa, P.\ Carruthers and
           B.\ Buschbeck,
           Phys.\ Rev.\ D {\bf 48}, 2040 (1993).

\bibitem{Egg93d}H.\ C.\ Eggers, P.\ Lipa, P.\ Carruthers
           and B.\ Buschbeck,
           Phys.\ Lett.\ {\bf 301B}, 298 (1993).

\bibitem{Lip91a}
           J.\ M.\ Alberty, R.\ Peschanski and A.\ Bia\l as,
           Z.\ Phys.\ {\bf C52}, 297 (1991);
           P.\ Lipa, H.C.\ Eggers, F.\ Botterweck and M.\ Charlet,
           Z.\ Phys.\ {\bf C54}, 115 (1991).


\bibitem{EMU01-92a}EMU01 Collaboration, M.\ I.\ Adamovich et al.,
              Nucl.\ Phys.\ B {\bf 388}, 3 (1992).

\bibitem{Lip90a}P.\ Lipa, U.\ of Vienna PhD dissertation (1990),
           (unpublished).

\bibitem{Lip91b}P.\ Lipa, in
               {\it Proceedings of the Ringberg Workshop on
                Multiparticle Production}, 1991,
               edited by R.\ C.\ Hwa, W.\ Ochs and N.\ Schmitz,
               World Scientific, (1992).

\bibitem{Kra93}H.\ C.\ Eggers and P.\ Lipa, in:
         {\it Soft Physics and Fluctuations}, Proceedings of the
         Cracow Workshop on Multiparticle Production, Krakow, Poland,
         3--8 May 1993 (to be published); see also Los Alamos
         Bulletin Board, hep-ph/9309256.

\bibitem{KS94}K.\ Kadija and P.\ Seyboth,
          Z.\ Phys.\ {\bf C61}, 465 (1994).

\bibitem{Lip92a}P.\ Lipa, P.\ Carruthers, H.\ C.\ Eggers and
           B.\ Buschbeck, Phys.\ Lett.\ {\bf 285B}, 300 (1992).

\bibitem{Car90d}P.\ Carruthers, H.C.\ Eggers, and I.\ Sarcevic,
           Phys.\ Lett.\ {\bf 254B}, 258 (1991);
         A.\ H.\ Mueller, Phys.\ Rev.\ D {\bf 4}, 150 (1977).

\bibitem{Col92a}P.\ H.\ Coleman and L.\ Pietronero,
           Phys.\ Rep.\ {\bf 213}, 311 (1992).

\bibitem{EMU01-93a}EMU01 Collaboration, M.\ I.\ Adamovich et al.,
              Phys.\ Rev.\ D {\bf 47}, 3726 (1993).

\bibitem{NA22-93a}EHS/NA22 Collaboration, N.\ M.\ Agababyan et al.,
           preprint HEN--353 (1993).

\bibitem{Pre89a}W.\ H.\ Press, B.\ P.\ Flannery, S.\ A.\ Teukolsky
          and W.\ T.\ Vetterling, {\it Numerical Recipes in Fortran},
           Cambridge University Press (1989).

\bibitem{Jur89a}I.\ Juricic {\it et al.},
           Phys.\ Rev.\ D {\bf 39}, 1 (1989);
            UA1 Collaboration, N.\ Neumeister {\it et al.},
            Phys.\ Lett.\ {\bf 275B}, 186 (1992).

\end{thebibliography}
\end{document}